\documentclass[12pt,preprint]{aastex}
\shorttitle{Pulsar Radio emission height}
\shortauthors{R. T. Gangadhara}
\received{}
\begin{document}
\title{PULSAR RADIO EMISSION ALTITUDE FROM CURVATURE RADIATION}
                                                                                
\author{R. T. Gangadhara}
                                                                                
\affil{Indian Institute of Astrophysics, Bangalore -- 560034, India\\
\email{ganga@iiap.res.in}}
%%%%%%%%%%%%%%%%%%%%%%%%%%%%%%%%%%%%%%%%%%%%%%%%%%%%%%%%%%%%%%%%
\begin{abstract}

        We assume that the relativistic sources moving along the dipolar magnetic 
field lines emit curvature radiation. The beamed emission occurs in the direction 
of tangents to the field lines, and to receive it, the sight line must align with the 
tangent within the beaming angle 
$1/\gamma,$ where $\gamma$ is the particle Lorentz factor. By solving the viewing 
geometry in an inclined and rotating dipole magnetic field, we show that, at any given 
pulse phase, observer tends to receive radiation only from the specific heights allowed by 
the geometry. We find outer conal components are emitted at higher altitudes compared 
to  inner components including the core. At any pulse phase, low frequency emission 
comes from  higher altitudes than high frequency emission. 
We have modeled the emission heights of pulse components of PSR B0329+54, and 
estimated field line curvature radii and particle Lorentz 
factors in the emission regions.
\end{abstract}

\section{INTRODUCTION}

	Pulsar radio emission is generally thought to be  coherent radiation from the 
relativistic plasma streaming along the open magnetic field lines. The emission beam geometry has 
been widely attempted to interpret in terms of emission in purely dipolar magnetic field 
(e.g., Radhakrishnan \& Cooke 1969; Goldreich \& Julian 1969; Sturrock  1971; Ruderman 
\& Sutherland 1975;  Michel 1982; Blaskiewicz et al. 1991).

Radhakrishnan and Cooke (1969) have proposed the rotating vector model (RVM) 
to explain the polarization-angle (PA) traverse (S-curve) of pulsar profiles. It has been 
fitted to average PA curves, and found to be quite successful (e.g., Lyne \& Manchester 1988).
The three chief assumptions (Hibschman \& Arons 2001) of RVM model are 
(i) the relativistic plasma flow is collimated by strong dipolar magnetic field lines, 
(ii) the observed radiation is beamed in the direction of tangents to the field lines, and 
(iii) the polarization angle is at a fixed angle to the curvature of field lines, as is the 
case of vacuum curvature radiation. Based on the assumptions (i) and (ii), we have attempted
to locate the radio emission region in pulsar magnetosphere from which a distant observer 
receives the radiation. 

Statistical analysis of the distribution of pulse components within the pulse window indicates 
that the emission beam is conal (e.g., Mitra \& Deshpande 1999; Kijak \& Gil 2002).
Rankin (1993) has indicated that  inner pulse components are emitted at lower altitudes 
than the outer ones. Gangadhara and Gupta (2001, hereafter GG01) have estimated the emission 
height of different radio pulse components of PSR~B0329+54, 
based on  the aberration-retardation phase shift. They have also found that the
inner components of pulse profiles originate from lower altitudes than the outer components. 
In the following paper Gupta and Gangadhara (2003) have then
estimated the emission heights of 6 more pulsars using the same technique.
Recently, the emission heights of all those pulsars have been re-estimated by Dyks, Rudak and 
Harding (2003, hereafter DRH03) by revising the aberration phase--shift relation given in GG01.
Also, there are other claims for core and cone emission altitudes being different
(e.g., Mitra \& Rankin 2002) and roughly the same (e.g., Gil 1991).

In this paper, we solve the viewing geometry for receiving the radio waves by a distant 
observer in \S~2. By assuming the particle Lorentz factors, we show in \S~2.2 that curvature
radiation predicts the emission heights which are comparable to the radio emission heights
in PSR~B0329+54.  In \S~3, by utilizing the properties of curvature radiation and the 
emission altitudes estimated from aberration--retardation phase shift (DRH03), we estimate 
the field line curvature radii and particle Lorentz factors in the emission region.

\section{EMISSION--BEAM GEOMETRY}

 Consider a magnetic dipole situated at the origin with magnetic dipole moment (${\bf \hat  m}$) 
aligned with the rotation axis $({\bf \hat  \Omega}) .$ In polar coordinates, the 
equation for a dipole--field line (Alfv\'en \& Falth\"ammar 1963) is $ r=r_{\rm e} \sin^2\theta, $
where  $\theta$ is the magnetic colatitude and $r$ the distance from the origin. The 
parameter $r_{\rm e}$ is the {\it field line constant},
which is a distance from origin to the point of field line intersection with the equatorial
plane $(\theta=\pi/2).$ In the case of an aligned dipole, $r_{\rm e}=r_{\rm LC}$ for the
last open field line, where $r_{\rm LC}=P c/2\pi$ is the light cylinder radius,
$c$ the speed of light, and $P$ the pulsar period.  In the  Cartesian coordinate system
with z-axis parallel to ${\bf \hat \Omega},$  the position vector of an 
arbitrary point Q on a field line is given by
\begin{equation}
{\bf r}_{\rm c} = r_{\rm e} ( \sin^3 \theta \cos\phi,\,  \sin^3 \theta \sin\phi,\,
 \sin^2\theta \cos\theta),
\end{equation}
where $\phi$ is the magnetic azimuth.

Now consider the situation where the dipole $({\bf \hat m})$ is 
{\it inclined} through an angle $\alpha$ with respect to ${\bf \hat \Omega}$, and
{\it rotated} by phase $\phi'$ around z-axis, as shown in Figure~1.
The position vector of the point Q on a magnetic field line, which is tilted and 
rotated, is given by
\begin{equation}\label{fleq}
{\bf r}_{\rm ct}={\Lambda} \cdot {\bf r}_{\rm c}\quad ,
\end{equation}
where the transformation matrix $ {\Lambda} = {\it R} \cdot {\it I} \, .$
The matrices for tilt (inclination) $\it I$ and rotation $\bf R$ 
are given by
\begin{eqnarray}
{\it I} = \left(
\begin{array}{ccc}
\cos\alpha &   0 & \sin\alpha\\
0 & 1 & 0 \\
-\sin\alpha & 0 & \cos\alpha \\
\end{array}  \right) 
\: ,\quad
{\it R} & = & \left(
\begin{array}{ccc}
\cos\phi ' & -\sin\phi ' & 0 \\
\sin\phi ' & \cos\phi ' & 0 \\
0 & 0 & 1
\end{array}     \right)\: .
\end{eqnarray}
The matrix $\it I$ produces clockwise rotation of the dipole around y-axis,
and $\it R$ counter-clockwise rotation around the z-axis.

At Q, we find the {\it tangent} to the field line by evaluating
$ {\bf b}_{\rm t} = {\partial{\bf r}_{\rm ct}}/{\partial \theta},  $ and
the {\it curvature}  by evaluating
$ {\bf k}_{\rm t}= {d {\bf \hat b}_{\rm t}}/{ds} \: , $
where  ${\bf \hat b}_{\rm t}
={\bf b}_{\rm t}/\vert {\bf b}_{\rm t}\vert $ is a unit tangent vector,
$ds$ is the arc length of the field line and $\vert{\bf b}_t\vert =
r_{\rm e}\,{\sqrt{5 + 3\,\cos (2\,\theta)}}\,\sin \theta/\sqrt{2}.$ Therefore, 
the field line curvature radius is given by
\begin{equation}\label{rho}
\rho=\frac{1}{\vert {\bf k}_{\rm t}\vert} =
\frac{r_{\rm e}\sin\theta\, [5 + 3 \cos(2\,\theta)]^{3/2}}{3\sqrt{2}\,
   [3 + \cos(2\,\theta)]}.
\end{equation}

Since ${\bf \hat m}$ is chosen to be parallel to ${\bf \hat z},$ the transformed magnetic 
dipole moment is given by
\begin{equation}
{\bf \hat m}_{\rm t} = {\Lambda }\, \cdot\, {\bf \hat z}=(\sin\alpha\,  \cos\phi ',\: 
\sin\alpha\,  \sin\phi ' ,\:  \cos\alpha ).
\end{equation}
Hence the magnetic field of a dipole (Jackson 1975), which is inclined and rotated, is 
given by
\begin{equation}
{\bf B}_{\rm t} = B_0\left(\frac{r_{NS}}{r}\right)^3 [3\,({\bf \hat r} \cdot 
{\bf \hat m}_{\rm t})\, {\bf \hat r}-{\bf \hat m}_{\rm t}]\: ,
\end{equation}
where $B_0$ is the surface magnetic field and $r_{\rm NS}\sim 10$~km is the neutron star 
radius. It can be easily shown that ${\bf B}_{\rm t}$ satisfies 
$\nabla \cdot {\bf B}_{\rm t}=0.$

  The dominant magnetic field in the emission region is often shown to be consistent
with being dipolar and to study the shape of pulsar radio beams (e.g., Narayan \& 
Vivekanand 1983; Lyne \& Manchester 1988; Kramer et al. 1997). In a static dipole magnetic 
field approximation, the basic features of the pulsar magnetosphere can be understood.
On the time scales of the order of pulse-phase bin, which is very small compared to the 
pulsar period, the rotating dipole may be treated, approximately, a static dipole.

\subsection{\it Magnetic Colatitude and Azimuth of Emission Spot}

Consider the sight line ${\bf \hat n}=(\sin\zeta ,\: 0,\:  \cos\zeta \:) ,$ 
which lies in the xz--plane and makes an angle 
$\zeta$ with respect to  ${\bf \hat \Omega},$
where $\zeta=\alpha+\beta,$ and $\beta$ is the angle of closest approach  of
sight line with respect to the magnetic axis. The half opening angle $(\Gamma)$ of 
the emission beam is given by
\begin{equation}\label{Gamma}
\cos\Gamma={\bf \hat n} \cdot {\bf \hat m}_{\rm t}=
\cos\alpha  \cos\zeta +   \sin\alpha \, \sin\zeta\, \cos\phi ' \: .
\end{equation}
If  $\tau$ is the angle between ${\bf \hat b}_{\rm t}$ and $
{\bf \hat m}_{\rm t}$ then we have 
\begin{equation}\label{tau}
\cos\tau = {\bf \hat b}_{\rm t}\cdot{\bf \hat m}_{\rm t} = \frac{1 + 3\,\cos (2\,\theta)}{{\sqrt{10 + 6\,\cos (2\,\theta)}}}\, .
\end{equation}
In relativistic flow, radiation is beamed in the direction of
field line tangent. So, at any instant, observable radiation comes from
a spot in the magnetosphere where the tangent vector ${\bf \hat b}_{\rm t}$ points
in the direction of ${\bf \hat n}.$ That is $\Gamma\approx\tau$ at the 
emission spot. Therefore, using equations~(\ref{Gamma}) and
(\ref{tau}), we can find the relation for {\it magnetic colatitude}
$\theta$ as a function of $\Gamma : $ 
\begin{equation}\label{theta}
\cos(2\,\theta)=\frac{1}{3}(\cos \Gamma\,\sqrt{8 + \cos^2 \Gamma}-\sin^2 \Gamma)\:, 
                   \quad\quad -\pi\leq\Gamma\leq\pi\: .
\end{equation}
This equation is similar to the one given in GG01 (see, eq.~4) for $\theta .$ However, 
the equation (\ref{theta}) is superior compared to GG01's equation as it
represents a single physical solution when $\Gamma$ changes sign. For $\Gamma\ll 1,$
equation~(\ref{theta}) reduces to the well known approximate form 
$\theta\approx \frac{2}{3}\Gamma .$

Next, let $\kappa$ be the angle between ${\bf \hat n}$ and ${\bf \hat b}_{\rm t},$ then we have
\begin{eqnarray}
\cos\kappa\!\!\! & =&\!\!\!{\bf \hat n}\cdot {\bf \hat b}_{\rm t}\nonumber\\
\!\!\!& = &\!\!\! \cos^2\Gamma
+\left(\cos \alpha\,\sin \zeta \,\cos \phi'\!-\sin \alpha\, \cos \zeta
      \right) \,\sin \Gamma \cos\phi -\sin \zeta\,\sin \phi'\sin \Gamma \sin\phi .  \label{kappa}
\end{eqnarray}
Again to receive the radiation $\kappa\sim 0,$ therefore,  we find the {\it magnetic azimuth} $\phi$ of the emission spot by solving ${\bf \hat n}\cdot {\bf \hat b}_{\rm t}=1$ and 
${\bf \hat n}\times {\bf\hat b}_t=0 :$
\begin{eqnarray}
\sin\phi\!\!\! &=&\!\!\! -\sin \zeta\,\sin \phi'\csc \Gamma ,\label{sphi}\nonumber\\
\cos\phi\!\!\! &=&\!\!\! (\cos\alpha \,\sin\zeta\,\cos\phi'-\cos\zeta\,\sin\alpha)\csc \Gamma .
\nonumber\label{cphi}
\end{eqnarray}
Therefore, we have
\begin{equation}\label{phi}
\phi=\arctan\left (\frac{\sin \zeta\,\sin \phi'}{\cos\zeta\,\sin\alpha-\cos\alpha 
\,\sin\zeta\,\cos\phi'}\right).
\end{equation}
Note that $\phi$ is measured with respect to the meridional plane defined by 
${\bf \hat \Omega}$ and ${\bf \hat m}_{\rm t}.$ 

\subsection{\it Radio Emission Height}

Pulsar radio emission is generally believed to be coherent curvature radiation by 
secondary pair plasma streaming along the dipolar magnetic field lines.
The characteristic frequency of curvature radiation, at
which the emission peaks, is given by (Ruderman \& Sutherland 1975):
\begin{equation}\label{freq}
\nu=\frac{3c}{4\pi}\frac{\gamma^3}{\rho},
\end{equation} 
where $\gamma$ and $\rho$ are the Lorentz factor and the radius of curvature of 
trajectory of a relativistic particle, respectively. Since particles closely follow 
the curved dipolar field lines, the curvature of particle trajectory can be 
approximated with the curvature of field lines. 

For the given $\nu$ and $\gamma$, equation~(\ref{freq}) predicts the value of $\rho$ 
required. Then using equation~(\ref{rho}) we can find the field 
line constant $r_{\rm e}$ with the help of $\theta_{\rm em}$  obtained from  
equation~(\ref{theta}). The azimuth $\phi_{\rm em}$ of the emission spot can be 
determined by using equation~(\ref{phi}). Hence, by having defined $(r_{\rm e}, 
\theta_{\rm em}, \phi_{\rm em}),$ we can estimate the emission height $r_{\rm em}$ 
of radio waves of frequency $\nu$ with respect to pulse phase $\phi'$
 using equation~(\ref{fleq}).

Based on the aberration-retardation phase-shift of pulse components in observed 
profiles, Gangadhara and Gupta (2001) have estimated the emission heights of pulse 
components in PSR~B0329+54 at frequencies 325~MHz and 606~MHz. Recently, 
Dyks, Rudak and Harding (2003) have revised the aberration phase--shift relation
given in GG01, and re-estimated the emission heights. In Figure~2, we have 
plotted the revised emission heights (marked points with error bars). 

For modeling the emission heights in PSR~B0329+54, we adopt $(\alpha, \beta)=(30^\circ, 2.1^\circ)$
given by Rankin (1993). We find that the curvature emission predicts the emission heights 
(solid and dashed line curves in Figure~2) which are comparable to the observed ones (DRH03) 
if the secondary particle Lorentz factor $\gamma\approx 340$ and 390 for the emissions at 
325~MHz and 606~MHz, respectively. Figure~2 shows that  
at any given frequency the emission near 
the pulse center (core) comes from lower heights compared to outer conal components, in 
agreement with the results derived from observations (marked points).
At any pulse phase, low frequency emission comes from higher altitude than   
the high frequency emission, in agreement with the well known radius--to--frequency 
mapping (RFM) (e.g., Cordes 1978; Phillips 1992). We find RFM mapping is  not uniform across the pulse window, i.e., it is more pronounced in the case of outer cons 
(say, cones 3 \& 4) than in the inner cones (say, cones 1 \& 2).
Mitra and Rankin (2002) have also
made a similar proposition from the study of 10 selected pulsars.
The emission height difference $\delta r_{\rm em}$ between the two frequencies
increase progressively with respect to pulse phase, and the
marked points in Figure~2 indicate
$\delta r_{\rm em}\approx  14 + 0.046\,\phi'\,^2 + 0.003\,\phi'\,^4$~km,
where the pulse phase $\phi'$ is in degrees.

\subsection{\it The Polar Cap}

 The polar cap boundary is defined by the last open field lines for which $ {\bf k}_{t} \cdot {\bf \hat z} =0$ at the light cylinder.
Therefore, the magnetic colatitude $\theta_{\rm lof}=\theta$, at which 
$ {\bf k}_{\rm t} \cdot{\bf \hat z}$ vanishes, is given by
\begin{equation}
\cos(2\,\theta_{\rm lof}) =\frac{-3\, {a_1}^2 - a_2\, \sqrt{8\, {a_1}^2 + {a_2}^2}}{9\, 
{a_1}^2 + {a_2}^2}\: , \quad\quad
-\frac{\pi}{2}<\phi\leq\frac{\pi}{2}
\end{equation}
where $a_1 = \sin\alpha\, \cos\phi$ and $a_2 = 3\, \cos\alpha.$ Next, for the range of  
$\pi/2<\phi\leq 3\pi/2,$  the colatitude is given by $\pi-\theta_{\rm lof}.$
The angle ($\eta$) between ${\bf \hat r}_{\rm ct}$ and ${\bf \hat z}$ is given by
\begin{equation}\label{eta}
\cos\eta = {\bf \hat z} \cdot {\bf \hat r}_{\rm ct} = \cos\alpha\, \cos\theta  -
 \sin\alpha\,  \sin\theta\, \cos\phi .
\end{equation}
If $\eta_{\rm lof}=\eta$  at the light
cylinder for last open field line then we have $\vert{\bf r}_{\rm ct}\vert \sin\eta_{\rm lof} 
= r_{\rm LC},$  and the last open field line constant
\begin{equation}
r_{\rm e, lof} = r_{\rm LC}\csc^2\theta_{\rm lof}\,\csc\eta_{\rm lof}\: .
\end{equation}
Hence the radial position of foot of a last open field line from the magnetic axis is
 given by
\begin{equation}
s_{\rm p} = r_{\rm NS} \theta_{\rm p}, 
\end{equation}
where $\theta_{\rm p} = \arcsin(\sqrt{r_{\rm NS}/r_{\rm e, lof}}) $
is the colatitude of foot of last open field line.

To plot polar cap consider the Cartesian coordinate system:
$({\rm x}_{\rm B},\, {\rm y}_{\rm B},\, {\rm z}_{\rm B})$ 
such that the axis ${\rm z}_{\rm B}$ is parallel to ${\bf \hat m}_{\rm t}$ and 
${\rm x}_{\rm B}$ lies in the meridional plane. The coordinates of foot of a last 
open field line is given by
\begin{equation}
(x_{\rm B}\:,y_{\rm B})  = (s_{\rm p}\cos\phi\:, s_{\rm p} \sin\phi). 
\end{equation}
Using $\alpha = 30^\circ,$ the polar cap of PSR~B0329+54 is plotted in Figure~3. 
It is nearly elliptical with the dimension of 164~m in the ${\rm x}_{\rm B}$ direction 
and 171~m in the direction of ${\rm y}_{\rm B},$ in agreement with the polar cap shapes 
proposed by Biggs (1990).

For a given pulse phase $\phi'$, using equations~(\ref{theta}) and (\ref{phi}), we can 
find $\theta_{\rm em}$ and $\phi_{\rm em}$ of the emission spot. The radial location of foot of 
field lines, from which a distant observer receives the radiation, is given by
\begin{equation}
s_{\rm em} = \sqrt{\frac{r_{\rm NS}^3}{r_{\rm em}}}\sin\theta_{\rm em}.
\end{equation}
Thus, by knowing  $s_{\rm em}$ and 
$\phi_{\rm em}$  one can specify $(x_{\rm B},\, y_{\rm B})$ of the foot of field lines, which 
direct the radiation beam towards the observer. The values of $s_{\rm em}$ estimated from the 
conal aberration-retardation phase-shift (marked points with error bars in $s_{\rm em}$) are 
plotted in Figure~3. The model (dashed and solid line) curves represent the locus of foot of 
those field lines, which are accessible to a distant observer in 325~MHz and 606~MHz observations, 
respectively. We find the low frequency emission is received from the field lines which lie closer 
to the magnetic axis than the high frequency ones. The curve (solid line) indicates increasing 
conal ring spacing $(\delta s_{\rm em})$ between the successive rings from inner 
(cone 1) to outer (cone 4) on polar cap: $\sim 14$~m between the 1$^{\rm st}$ ring and the 
2$^{\rm st}$ ring, $16$~m between 2$^{\rm nd}$  and 3$^{\rm rd}$, and $19$~m between 
3$^{\rm rd}$ and 4$^{\rm th}.$ The dashed line curve for 325~MHz emission also indicate the 
similar ring spacings, except they are less by $\sim 0.5$~m. These results tend to support 
the model of  concentric rings of sparks produced in the vacuum gap region just above the 
neutron surface (e.g., Ruderman \& Sutherland 1975; Gil \& Sendyk 2000).

\section{DISCUSSION}

In the previous section, to explain the emission height of components in PSR~B0329+54,
we selected the Lorentz factor $\gamma$ of about 340 and 390 for the emissions at 325~MHz 
and 606~MHz, respectively. It means  $\rho$ and $\gamma$ assume constant values across the 
pulse window. However, it may not be reality, as the model heights significantly depart 
from the observed ones (marked points) as indicated by Figure~2.   
On other hand, one can turn around the calculation by accepting the observed emission 
heights ask for the required $\gamma$ and $\rho .$ In column 2 of Tables 1 \& 2, we have 
given the phase location of conal components in the absence of 
aberration-retardation phase shift (see, eq.~11, GG01), i.e., in corotating frame. Next, 
the columns 3 and 4 give the
magnetic colatitude $\theta$ and the longitude $\phi$ of emission spot, estimated 
using equations~(\ref{theta}) and (\ref{phi}). 
The small values of $\theta < 6^\circ$ implies  the conal emission mostly comes from the 
vicinity of magnetic axis. By having known $r_{\rm em}$ and 
$\theta_{\rm em},$ we estimated the radius of curvature $\rho $ using equation~(\ref{rho}), and 
given in column~5.  It shows inner cones are emitted from the region of smaller curvature 
than outer ones. Next, for a given frequency $\nu$ and known
$\rho,$  equation~(\ref{freq}) allows to estimate the $\gamma$ of secondary particles, 
as given in column~6. It shows particles with slightly higher $\gamma$ move on 
the field lines associated with outer cones than inner ones.

The emission beam by relativistic particles is centered on their velocity,
which is roughly parallel to the field line tangent ${\bf \hat b}_{\rm t},$ and has a
radial angular width of $1/\gamma.$ Consequently, observer tends to receive the 
radiation even when ${\bf \hat n}$ does not perfectly align with ${\bf \hat b}_{\rm t},$
which can lead to a spread in the emission height as well as in pulse phase of component
peak location. So, we estimated the spread in emission height of conal components by allowing $\Gamma$ in equation~(\ref{theta}) to vary by $1/\gamma.$ However, we find the spread is minimal and lies within the error
bars. For example, at 325~MHz, we find 9~km for the $1^{\rm st}$ cone and 15~km for 
the $4^{\rm th}$ cone. At 606~MHz, it is 6~km for the $1^{\rm st}$ cone
and 10~km for the $4^{\rm th}.$ For other cones the spread lies in between 
these values. We also estimated the spread in phase of component peaks, and found to be less 
than $0.4^\circ$ for all the components.
If we compare the emission heights (DRH03) along with these spreads, we find
the emission regions of conal components at the two frequencies are well separated:
it is 14~km in the case of $1^{\rm st}$ and 214 km for the $4^{\rm th}.$ Further,
the pulse phase locations of conal components are different at the two frequencies,
as indicated by column 2 in Tables 1 \& 2. So, the observer tends to see the components 
separated in phase as well. Note that the received radiation is maximum only when 
${\bf \hat n}$ perfectly aligns with ${\bf \hat b}_{\rm t}.$ Hence we may conclude that the
observer tends to receive the radiation at the two frequencies from two distinctly located 
regions in the magnetosphere. 
 
  Using these values of $\gamma$ and $\rho,$ we estimated the power radiated by 
an electron or a positron by curvature emission (see, eq.~25a,  Ruderman \& Sutherland 1975),
and given in column~7. It shows the single particle emission is  maximum in the 
region of inner cone than in the outer cone regions. However, we know from the 
estimates made by Ruderman and Sutherland (1975) that the incoherent 
superposition of single particle radiation cannot explain the pulsar radio emission.
The very high brightness temperature ($10^{25}$--$10^{30}$~K) of pulsars lead to the 
conclusion that it must be coherent. The coherence due to the bunching 
of plasma particles in the magnetosphere was proposed much earlier (e.g., 
Pacini \& Rees 1970; Sturrock 1971). The coherence due to bunching goes as
inverse of wavelength of radio waves, therefore,  the emission by bunches 
at low frequency tends to dominate the high frequency emission. 

Thus, we have developed a method for estimating the possible values of $\gamma$ and 
$\rho$ needed for the emission of radio waves at a given frequency.  By matching 
the coherent curvature emission with the observed pulsar fluxes, one may be able to 
estimate the plasma density and the coherency (bunching) factor exist in the
radio emission region of pulsars. For $\nu\geq\nu_p$ strong coherent radiation occurs, 
where $\nu_{\rm p}$ is the plasma frequency (Ruderman \& Sutherland 1975).

While estimating the aberration-retardation phase shift of conal components, 
the core (reference) phase was set to zero (GG01; DRH03). However, it 
is possible that the core is also emitted at a finite height from the neutron star surface, 
as indicated by Figure~2. We fitted a 4$^{\rm th}$ degree polynomial to the emission
heights (marked points in Fig.~2) of 8 conal components. At zero phase, the fit 
indicated emission height of $26\pm50$~km for the 325~MHz points and $12 \pm 45$~km for the 606~MHz points.
These heights may be treated as minimum values for the core emission height, as the 
component heights themselves have been estimated by setting the core phase to zero. Though 
it is important to consider the emission height of core, we believe our results will not 
change significantly by the core height, as aberration and retardation are minimal at 
that height.

\section{CONCLUSION}

By assuming the curvature emission by relativistic sources is beamed in the direction of tangent
to the dipolar magnetic field lines, we have resolved the pulsar radio 
emission geometry by solving the viewing geometry in an inclined and rotating
dipolar magnetic field.  Due to the  geometric 
restrictions,  a distant observer tends to receive outer conal components from  higher
altitudes than inner ones including the core. Further, we find low frequency emission 
comes from higher altitudes than high frequency emission, in agreement with the 
radius-to-frequency mapping (RFM). We find RFM is more pronounced 
in the outer conal components than in the inner ones.
Using the emission heights obtained from aberration-retardation phase shift,
we have estimated the probable values of Lorentz factor and field line curvature radii in the 
emission region.
\begin{acknowledgements}
I would like to thank Don Melrose and Simon Johnston for illuminating discussions, and
J. M. Rankin, J. A. Gil, Y. Gupta, V. Krishan, and C. Sivaram for comments. I thank
anonymous referee for useful comments.
\end{acknowledgements}
%----------------------------------------------------------------------------------

\clearpage
\begin{table}
\begin{center}
\caption{
Parameters related to radio emission from PSR~B0329$+$54 at 325~MHz
\label{tbl_1}}
\begin{tabular}{crrrrrr}
\tableline\tableline
Cone &\multicolumn{1}{c}{$\phi'$} &\multicolumn{1}{c} {$\theta$} &\multicolumn{1}
{c}{$\phi$} &\multicolumn{1}{c} {$\rho/r_{\rm LC}$} &
\multicolumn{1}{c}{$\gamma$}   & \multicolumn{1}{c}{$P/P_1$\tablenotemark{\dag}} \\
No.  &\multicolumn{1}{c}{(deg)} &\multicolumn{1}{c} {(deg)} &\multicolumn{1}
{c}{(deg)} &  & &                   \\
\tableline
1 ...... & 5.0 $\pm$ 0.26 & 2.22 $\pm$ 0.07 & $-$53.0 $\pm$ 1.6 &
   0.15 $\pm$ 0.00 & 286 $\pm$ 3 & 1.00 $\pm$ 0.08 \\
2 ......& 8.4 $\pm$ 0.19 & 3.21 $\pm$ 0.06 & $-$67.8 $\pm$ 0.6 &
  0.23 $\pm$ 0.00 & 329 $\pm$ 2 & 0.76 $\pm$ 0.03 \\
3 ......& 12.0 $\pm$ 0.27 & 4.35 $\pm$ 0.09 & $-$76.5 $\pm$ 0.5 &
  0.31 $\pm$ 0.01 & 363 $\pm$ 2 & 0.62 $\pm$ 0.03 \\
4 ......& 17.0 $\pm$ 0.80 & 5.99 $\pm$ 0.26 & $-$83.9 $\pm$ 1.0 &
  0.33 $\pm$ 0.01 & 370 $\pm$ 5 & 0.60 $\pm$ 0.06 \\
\tableline
\\
\multicolumn{3}{l}{{\footnotesize  $^{\rm \dag}$ 
                  $P_1=1.16\times 10^{-16}$ erg s$^{-1}$}} \\
\end{tabular}
\end{center}
\end{table}

\begin{table}
\begin{center}
\caption{Parameters related to radio emission from PSR~B0329$+$54 at 606~MHz
\label{tbl_2}}
\begin{tabular}{crrrrrr}
\tableline\tableline
Cone &\multicolumn{1}{c}{$\phi'$} &\multicolumn{1}{c} {$\theta$} &\multicolumn{1}
{c}{$\phi$} &\multicolumn{1}{c} {$\rho/r_{\rm LC}$} &
\multicolumn{1}{c}{$\gamma$}   & \multicolumn{1}{c}{$P/P_1$\tablenotemark{\dag}} \\
No.  &\multicolumn{1}{c}{(deg)} &\multicolumn{1}{c} {(deg)} &\multicolumn{1}
{c}{(deg)} &  & &                   \\
\tableline
1 ......& 4.9 $\pm$ 0.25 & 2.19 $\pm$ 0.07 & $-$52.4 $\pm$ 1.6 & 0.12 $\pm$ 0.00 &
       328 $\pm$ 3 & 1.00 $\pm$ 0.07 \\
2 ......& 7.8 $\pm$ 0.18 & 3.02 $\pm$ 0.06 & $-$65.8 $\pm$ 0.6 & 0.21 $\pm$ 0.00
        & 391 $\pm$ 2 & 0.70 $\pm$ 0.03 \\
3 ......& 11.0 $\pm$ 0.45 & 4.02 $\pm$ 0.14 & $-$74.5 $\pm$ 1.0 & 0.26 $\pm$ 0.01
        & 419 $\pm$ 5 & 0.61 $\pm$ 0.05 \\
4 ......& 15.7 $\pm$ 0.59 & 5.56 $\pm$ 0.19 & $-$82.3 $\pm$ 0.8 & 0.26 $\pm$ 0.01
        & 420 $\pm$ 5 & 0.61 $\pm$ 0.05 \\
\tableline
\\
\multicolumn{3}{l}{{\footnotesize  $^{\rm \dag}$ 
                  $P_1=3.06\times 10^{-16}$ erg s$^{-1}$}} \\
\end{tabular}
\end{center}
\end{table}
\begin{figure}
\plotone{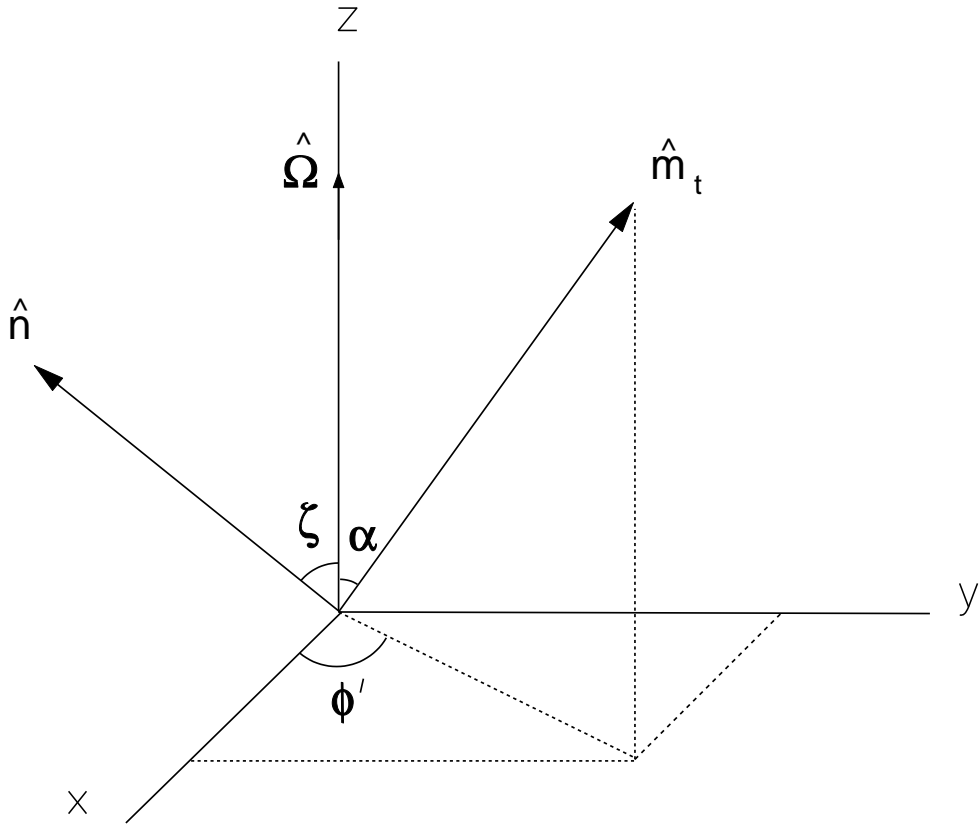}
\caption{Coordinate system to describe the inclined and
rotating magnetic dipole. The rotation phase $\phi'$ is measured from
the fiducial plane (x-z) in the counter-clockwise direction around ${\bf \hat \Omega}$. The magnetic colatitude $\theta$ is measured from ${\bf \hat m}_{\rm t},$ and the magnetic azimuth $\phi$ measured counter-clockwise around ${\bf \hat m}_{\rm t}$ from the
meridional plane defined by ${\bf \hat \Omega}$ and ${\bf \hat m}_{\rm t}$. }
\end{figure}
\begin{figure}
\plotone{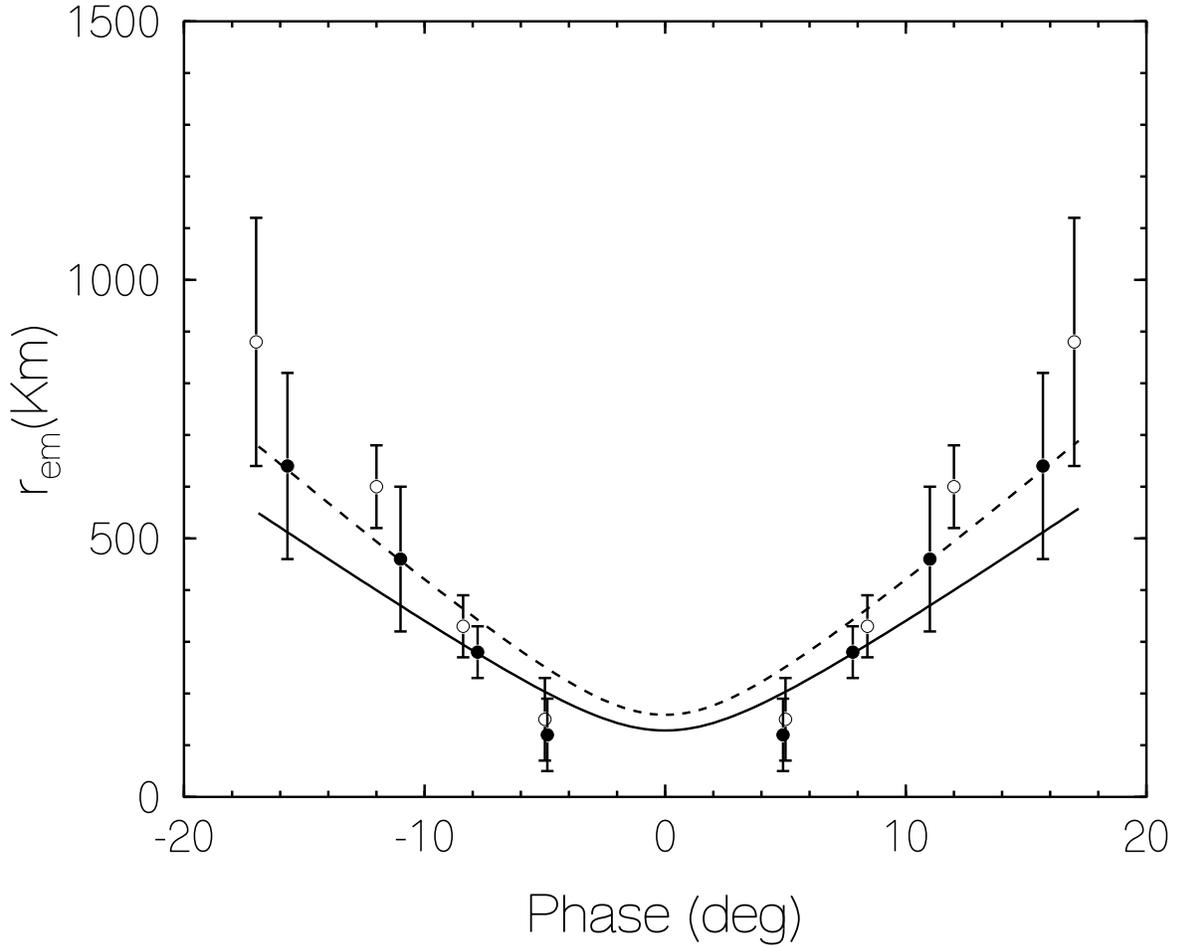}
\caption{Radio wave emission heights in PSR~B0329+54 with respect to rotation phase $\phi'$:
solid and dashed line curves are for the emissions at 606~MHz and 325~MHz, respectively. 
The emission heights estimated from aberration-retardation phase shift (DRH03) are 
superposed for comparison: the points marked with $\circ$ and {\large$\bullet$} are 
for the components at 325~MHz and 606~MHz, respectively.  }
\end{figure}
\begin{figure}
\plotone{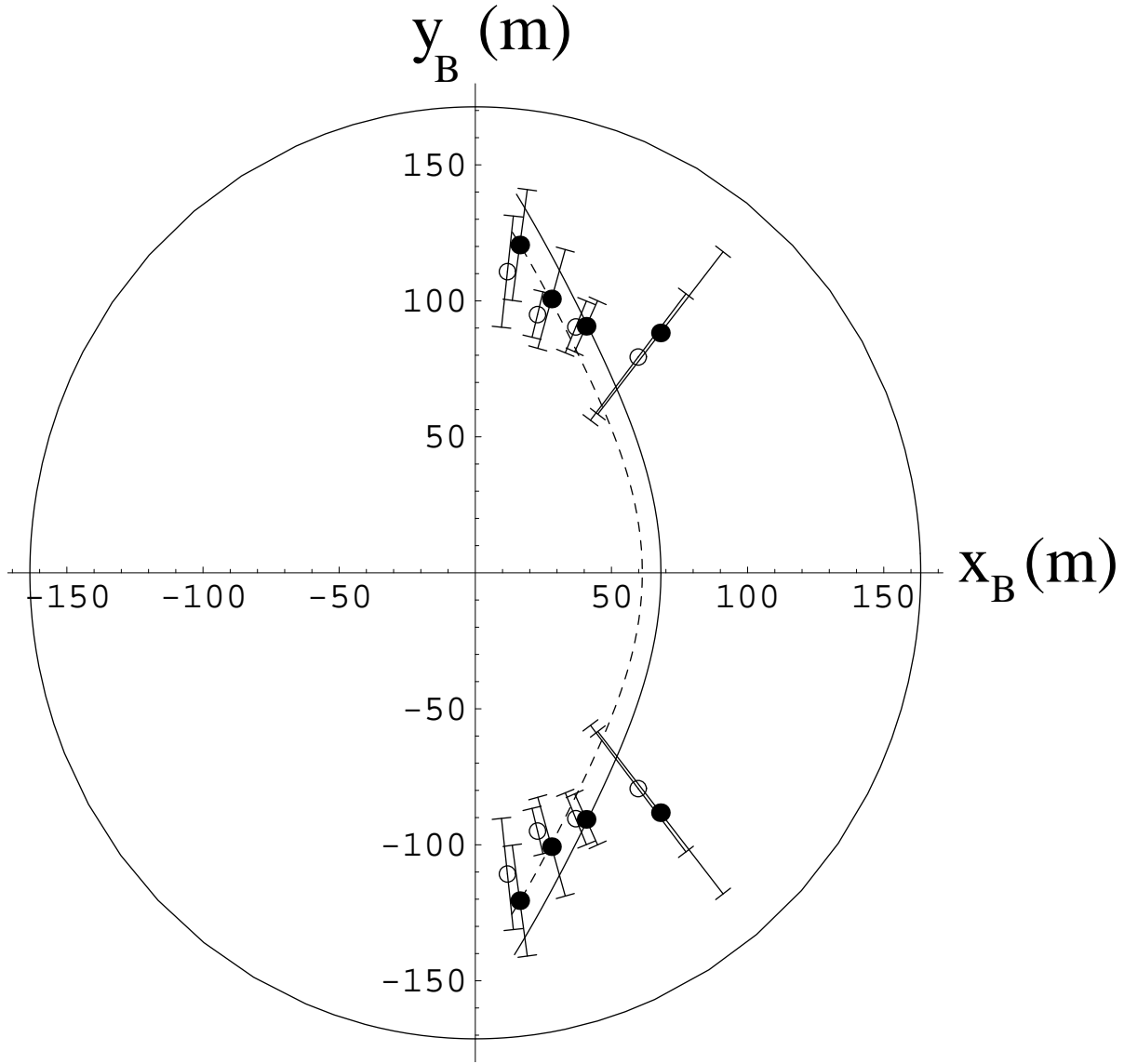}
\caption{Polar cap of PSR~B0329+54. Solid and dashed line curves represent the locus 
of foot of field lines associated with the emissions at 606~MHz and 325~MHz, respectively. 
The values of $s_{\rm em}$ estimated from aberration-retardation phase shift 
are superposed for comparison: the points marked with 
$\circ$ and {\large $\bullet$} are for 325~MHz and 606~MHz, 
respectively. }
\end{figure}
\end{document}